\documentclass[prl,aps,nofootinbib,superscriptaddress,twocolumn]{revtex4}
\usepackage{graphics}
\usepackage{amsmath}
\usepackage{epsfig}
\usepackage{wasysym}

\begin{document}

\title{Good NEWS for GeV Dark Matter Searches}

\author{Stefano Profumo}
\email{profumo@ucsc.edu}\affiliation{Department of Physics, University of California, 1156 High St., Santa Cruz, CA 95064, USA}\affiliation{Santa Cruz Institute for Particle Physics, Santa Cruz, CA 95064, USA}

\date{\today}

\begin{abstract}
\noindent The proposed NEWS apparatus, a spherical detector with a small central electrode sensor operating as a proportional counter, promises to explore new swaths of the direct detection parameter space in the GeV and sub-GeV Dark Matter particle mass range by employing very light nuclear targets, such as H and He, and by taking advantage of a very low (sub-keV) energy threshold. Here we discuss and study two example classes of Dark Matter models that will be tested with NEWS: GeV-scale millicharged Dark Matter, and a GeV-Dirac Fermion Dark Matter model with a light (MeV-GeV) scalar or vector mediator, and indicate the physical regions of parameter space the experiment can probe. 
\end{abstract}

\maketitle

\section{Introduction}
Large-volume spherical gas detectors possess an array of interesting features with broad applicability to a variety of scientific contexts, from radon emanation monitoring, to neutron flux counting, to the detection of low-energy neutrinos and low-mass weakly-interacting massive particles (WIMPs) \cite{Giomataris:2008ap, Dastgheibi-Fard:2014maa}. The detector concept and design are simple: a large spherical gas volume with a small central electrode sensor, forming a proportional counter. The central electrode is supported by a metallic rod, and is kept at high voltage. Electrons drift to the central sensor through low-field regions, eventually triggering an electron avalanche close to the sensor, where the electric field (varying as $1/r^2$) dramatically increases \cite{Giomataris:2008ap, Gerbier:2014jwa}. 

Spherical detectors offer low (sub-keV) energy thresholds, good energy resolution, single ionization electron sensitivity, and significant flexibility in the choice of the nature of the target gas and its pressure \cite{Gerbier:2014jwa, Dastgheibi-Fard:2014maa}. Studies of detector response and calibrations, focusing especially on the question of the optimal central spherical sensor size and geometry, are actively under way within the NEWS (New Experiments With Sphere)-SNO Collaboration\footnote{\tt https://www.snolab.ca/news-projects/index.html} \cite{Dastgheibi-Fard:2014maa}. 
 
The redundancy of target gas nuclei and the low-energy threshold achievable with the NEWS detector make this concept quite appealing, especially for low-mass particle Dark Matter searches, where such an apparatus would offer a highly complementary sensitivity to that of ongoing ton-scale noble-gas detectors \cite{Cushman:2013zza}. Preliminary estimates of the detector performance indeed indicate interesting prospects for using a NEWS-type detector in searching for GeV or even sub-GeV WIMPs \cite{Gerbier:2014jwa, GerbierPC}.

Here, we discuss two classes of GeV-mass Dark Matter models that are natural targets for NEWS: (i) millicharged Dark Matter with masses between 0.1 and 10 GeV and electric charge $10^{-6}$ to $10^{-9}$ times the electron electric charge $e$, and (ii) a Dirac fermion Dark Matter candidate of similar mass, but coupled to a light, MeV-GeV-scale vector or scalar mediator. For both classes of example models (which by no means exhaust the theory space accessible by the experiment under consideration here) we highlight the scientific potential of the NEWS detector concept and provide details on the portion of parameter space that NEWS will probe.

\section{Millicharged GeV Dark Matter}

The possibility that Dark Matter (DM) possesses an electromagnetic charge has been entertained in a variety of different contexts \cite{Goldberg:1986nk, DeRujula:1989fe, Dimopoulos:1989hk, Davidson:1993sj}. While strong constraints exist on DM particles with integer electric charge $\pm e$ \cite{Davidson:1993sj, Davidson:2000hf}, there are many theoretical possibilities for the effective charge of the DM to be a small fraction of the electron's charge, $q=\varepsilon e$, for example Stueckelberg models \cite{Feldman:2007wj} or models with a massless dark photon \cite{Holdom:1985ag}. In a generic way, any theory of ``millicharged'' DM is qualitatively described by the values of two parameters: the DM particle mass $m_X$, and the charge $\varepsilon$. We will hereafter adopt this model-independent description.

\begin{figure*}%
\includegraphics[width=1.5\columnwidth]{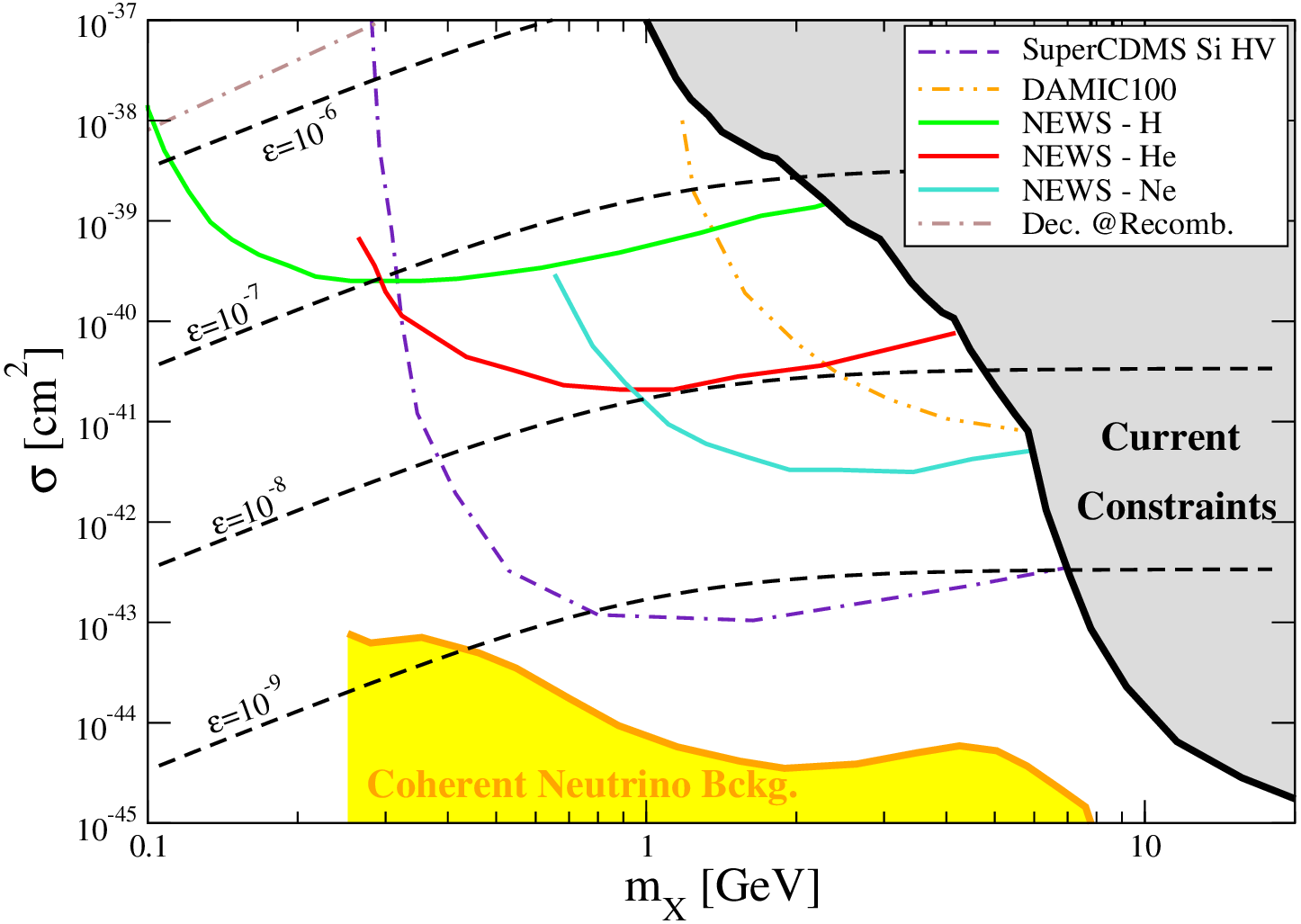}%
\caption{Spin-independent cross section of millicharged GeV Dark Matter off of protons; the grey region is ruled out by current experimental constraints from a variety of experiments \cite{Cushman:2013zza}; the yellow region indicates signal levels below coherent neutrino scattering \cite{Billard:2013qya}; the green, red and blue regions show the projected sensitivity for NEWS with H, He and Ne target gas \cite{Gerbier:2014jwa, GerbierPC}. For comparison, we also indicate the projected sensitivity of SuperCDMS Si HV \cite{supercdms} and of DAMIC100 \cite{Aguilar-Arevalo:2013uua}. Finally, the black dashed lines indicate points at constant millicharge $\varepsilon$ in units of $e$, and the dot-dashed brown line in the top left delimits (from below) the region where the Dark Matter would interfere with CMB recombination \cite{McDermott:2010pa}. }%
\label{fig:millicharged}%
\end{figure*}

Constraints on the $(m_X,\varepsilon)$ plane from cosmology and structure formation have been recently reviewed and updated in Ref.~\cite{McDermott:2010pa}, which also calculated rates for direct millicharged DM detection. Constraints from CMB data and baryon acoustic oscillations arise from requiring the DM to be decoupled from the baryon-photon plasma at recombination; such constraints effectively rule out the region of parameter space where millicharged DM can be produced as a thermal relic, thus implying either non-thermal production, or additional interactions besides those mediated by the millicharge \cite{McDermott:2010pa}. We indicate the constraints on the $(m_X,\varepsilon)$ plane with the brown dot-dashed line in the top left.

A particularly significant constraint on millicharged DM models in the region of parameter space relevant for direct detection stems from the shielding effect of the large-scale Milky Way magnetic field (first discussed in Ref.~\cite{Chuzhoy:2008zy}) and from the evacuation of particles from the Galactic plane due to supernova explosions \cite{McDermott:2010pa}.

Two important caveats to the constraints from shielding and evacuation of millicharged particles from the Galactic plane have been however presented in Ref.~\cite{Foot:2010yz}. First, in the presence of a ``paraphoton'', DM self-interactions randomize particle trajectories on scales much shorter than the gyro-radius; Secondly, Ref.~\cite{Foot:2010yz} points out that the millicharged DM plasma would include ``mirror'' electric and magnetic fields which would likely dominate the dynamics of DM particles over the effects of the Galactic magnetic field. On the parameter space we focus on, Ref.~\cite{Foot:2010yz} argues that the shielding/evacuation effects described in \cite{Chuzhoy:2008zy, McDermott:2010pa} are suppressed and therefore the resulting constraints do not apply. Quantitatively, one should ascertain that the typical distance scale needed to randomize the mirror particles' directions be much smaller than the mirror particles' gyroradius. Should this condition not hold, generically one should expect a suppression of the local dark matter density due to particle evacuation by the Galactic magnetic fields. Inputting the relevant model parameters into Eq.~(10) in Ref.~\cite{Foot:2010yz} one finds that only for $\varepsilon\lesssim 10^{-8}$ are trajectories randomized at much smaller scales than the gyroradius; however, the mirror particle plasma magnetic field can be much more significant to the particles' trajectories than the effects induced by the Galactic magnetic field, so larger values of $\varepsilon$ might also potentially not evacuate the Galactic disk. It is important to note that this conclusion does not apply universally to any millicharged DM theory, but that, rather, it is specific to models of ``mirror'' DM that contain a ``paraphoton''.

Here, we calculate the regions of the $(m_X,\varepsilon)$ parameter space that will be uniquely probed by the NEWS apparatus \cite{Gerbier:2014jwa}. To this end, we calculate and plot in fig.~\ref{fig:millicharged} the DM-proton spin-independent scattering cross section for a given $(m_X,\varepsilon)$ parameter space point, and compare with the anticipated sensitivity reach of NEWS for three different gas targets: H, He and Ne \cite{Gerbier:2014jwa}. We use a standard procedure to calculate the scattering rate (see e.g. sec. VI.A of Ref.~\cite{McDermott:2010pa}), which takes into account the different velocity dependence of the standard spin-independent case from the present scattering cross section off of nuclei ($\sigma\propto 1/q^4$, with $q$ the momentum transfer), and we translate the $(m,\varepsilon)$  predictions onto the $(m,\sigma)$ plane.

\begin{figure*}%
\includegraphics[width=1.5\columnwidth]{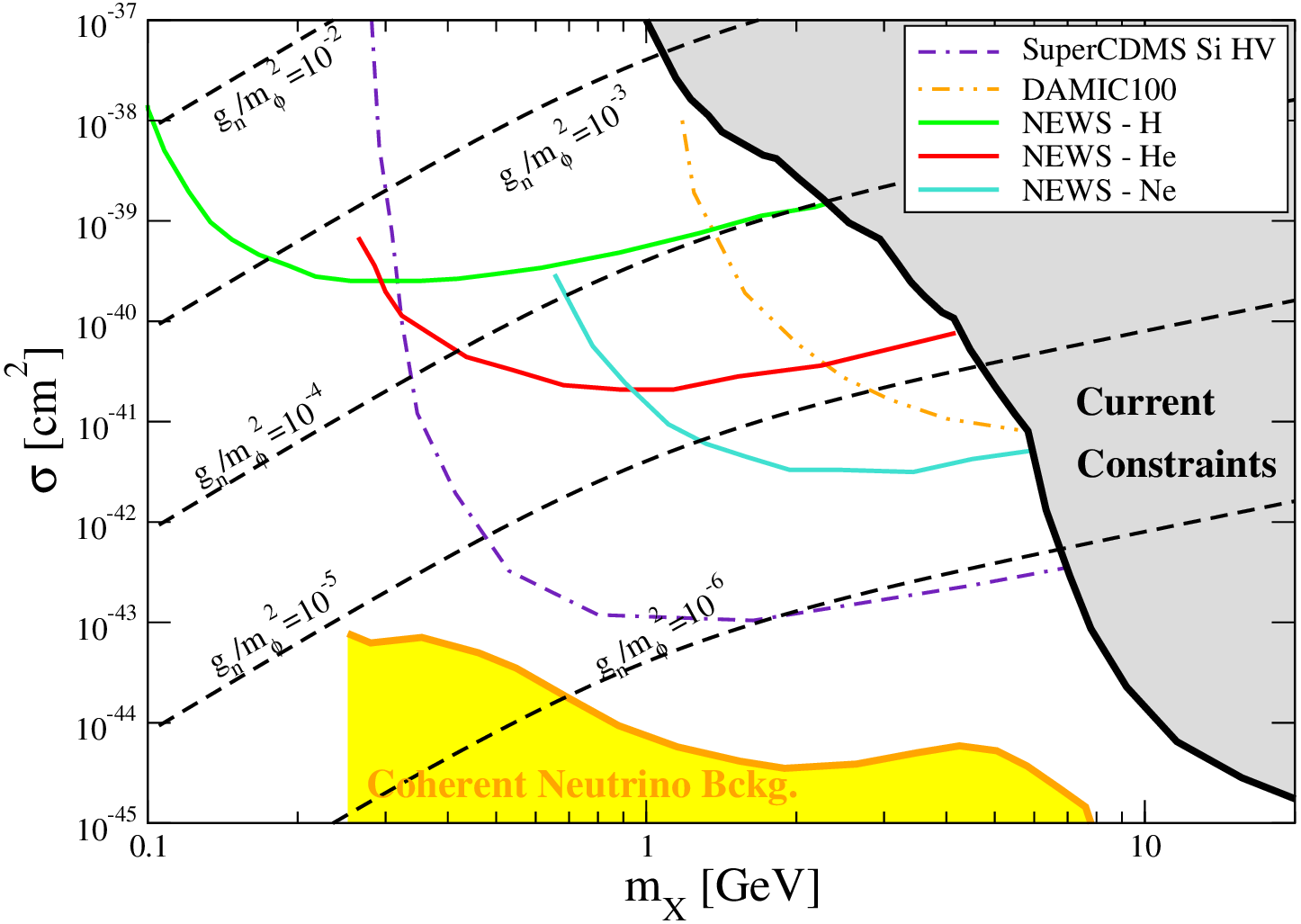}%
\caption{Spin-independent cross section of GeV Dark Matter with a light mediator off of protons; the conventions are the same as in fig.~\ref{fig:millicharged}, but the dashed black lines now indicate points at constant $g_n/m_\phi^2$ in units of ${\rm GeV}^{-2}$. Notice that to avoid constraints from CMB data a large degree of asymmetry between dark matter and anti-dark matter particles is required at low values for the dark matter mass; this implies, in turn, a model-dependent rescaling of the coupling $g_n$ (see the text for more details).}%
\label{fig:lightdm}%
\end{figure*}

In the figure we
show newly calculated  limits obtained
for around 100 kg$\times$day exposure with Ne/He/CH$_4$, taking
into account anticipated background from materials,
with threshold set at 1 electron (i.e. 20 to 40 evee) and
quenching factors extrapolated down to 100-200 evNR \cite{GerbierPC}.
For reference, we also show (in the upper left corner) the limit corresponding to requiring kinetic decoupling at or before the time of recombination (the excluded region is above the plotted line). The grey shaded region corresponds to a combination of current constraints including results from LUX, SuperCDMS LT, CDMSlite, and DAMIC \cite{Cushman:2013zza}. We also indicate the coherent neutrino scattering ``floor'' with the orange line \cite{Billard:2013qya}. Finally, for comparison, we show the projected sensitivity of SuperCDMS Si HV \cite{supercdms} with a double-dash-dotted purple line, and of the future DAMIC100 \cite{Aguilar-Arevalo:2013uua} with a double-dotted-dashed orange line.

Our results indicate that millicharged DM candidates with a mass between 0.15 GeV and 6 GeV and with millicharges between $2\times10^{-9}\lesssim\varepsilon\lesssim2\times10^{-6}$ are squarely within the anticipated NEWS sensitivity. We note that the very low-mass range (specifically, below approximately 0.3 GeV), with $10^{-7}\lesssim\varepsilon\lesssim\times10^{-6}$, will also be out of reach for the planned SuperCDMS Si HV sensitivity \cite{Sander:2012nia}, and will thus be uniquely testable with NEWS.

\section{GeV Dark Matter with a light Mediator}

We consider now theories where the DM is a Dirac fermion of mass $m_X$ coupled to a light scalar ($\phi$) or vector ($\phi_\mu$)  mediator of mass $m_\phi<m_X$. The relevant terms in the Lagrangian are, for the scalar and vector mediator cases, respectively, \cite{Lin:2011gj}
\begin{equation}
{\cal L}_S=g_X\bar X X\phi+g_f\bar f f\phi+m_X\bar X X+m_\phi^2\phi^2,
\end{equation}
\begin{equation}
{\cal L}_V=g_X\bar X \gamma^\mu X\phi_\mu+g_f\bar f \gamma^\mu f\phi_\mu+m_X\bar X X+m_\phi^2\phi^\mu\phi_\mu,
\end{equation}
where $f$ indicates a generic Standard Model fermion. In order to reduce the dimensionality of the parameter space, we fix $g_X$ to a value that would make the DM a thermal relic producing precisely the right number density of relics to explain the observed universal Dark Matter density; this requires setting $\alpha_X\equiv g_X^2/4\pi\sim 5.2\times 10^{-5}(m_X/{\rm GeV})$ \cite{Lin:2011gj}. The DM-proton spin-independent scattering cross section is then defined once, for example, the ratio of the coupling to nucleons $g_n=3\times g_q$ divided by the mass of the mediator squared, $g_n/m_\phi^2$, is given.

Pair-annihilation into light mediators and the subsequent decay into light Standard Model fermions distorts the CMB anisotropies and is thus constrained by CMB data. The constraints depend on details of the light mediator physics, for example whether the light mediator contributes to the number of effective light degrees of freedom, and whether Sommerfeld enhancement effects are present. However, in general, the asymptotic relative dark matter to anti-dark matter abundance $r_\infty\equiv \rho_{\bar X}/\rho_X$ is constrained as a function of mass to be much smaller than 1 for small $m_X$. In particular, the results of Ref.~\cite{Lin:2011gj} indicate that $r_\infty\lesssim 10^{-1}$ for $m_X\sim 10$ GeV, $r_\infty\lesssim 10^{-2}$ for $m_X\sim 1$ GeV, and $r_\infty\lesssim 10^{-3}$ for $m_X\sim 0.1$ GeV. In the very light dark matter regime, CMB constraints thus force a large degree of dark matter asymmetry. This, in turn, would require a larger value for $g_X$ to deplete the symmetric component (according for example to what shown in Fig.~2 of Ref.~\cite{Lin:2011gj}). The relevant value of $g_n$ would then be trivially re-scaled accordingly.

We illustrate our results in fig.~\ref{fig:lightdm}, which uses the same conventions as fig.~\ref{fig:millicharged} above, but where the dashed lines now indicate constant values for the ratio $g_n/m_\phi^2$. We conclude that NEWS will be key in probing the range $2\times10^{-6}\lesssim g_n/(m_\phi/{\rm GeV})^2\lesssim 10^{-2}$ over the range of masses $0.15\lesssim m_X/{\rm GeV}\lesssim 6$. As above, we note that in the low mass range, and for $g_n/(m_\phi/{\rm GeV})^2\sim {\rm few}\times 10^{-3}$ NEWS will be the only experiment able to test this rather generic DM setup.

\begin{figure}%
\includegraphics[width=\columnwidth]{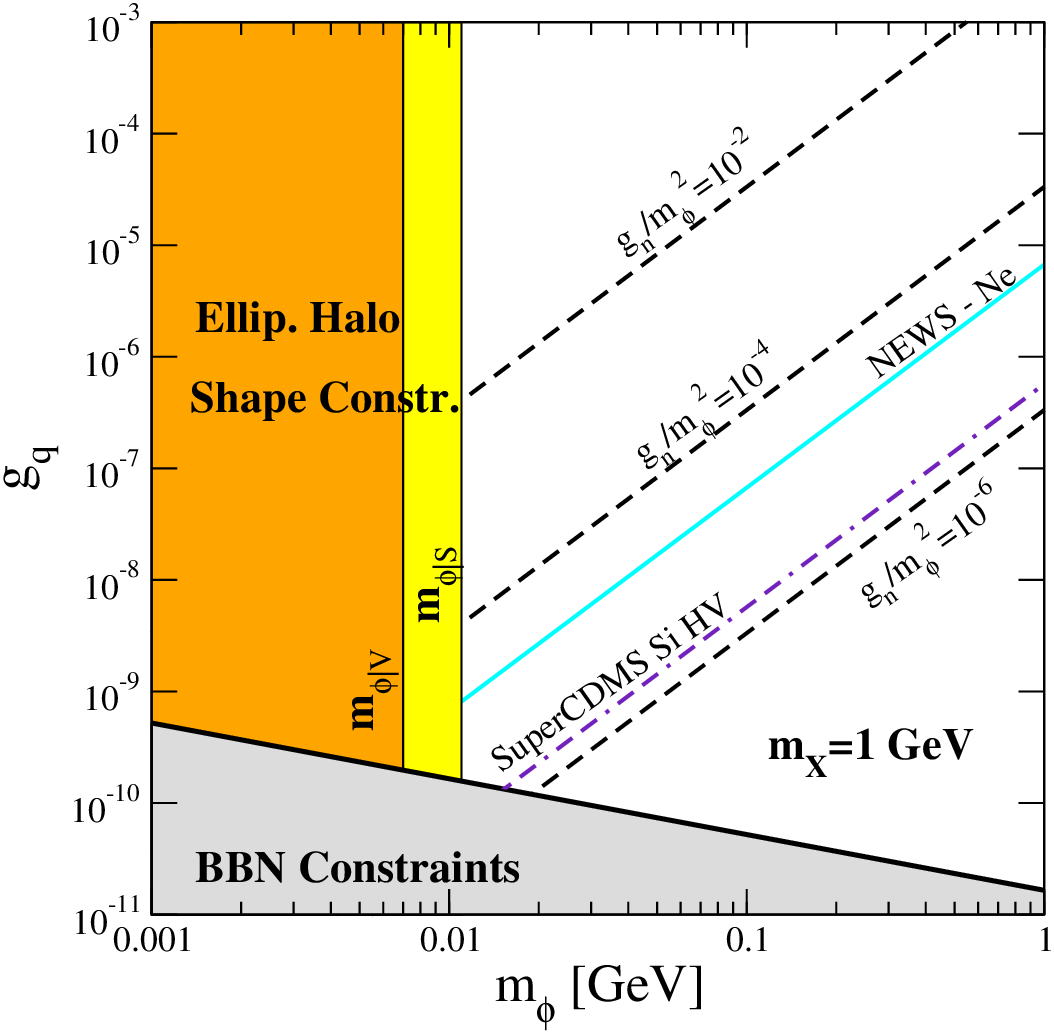}%
\caption{The parameter space for the mediator mass $m_\phi$ and coupling to quarks $g_q=g_n/3$, for a 1 GeV Dark Matter mass $m_X$; the lower left region shaded in grey has mediator lifetimes that exceed the epoch of Big Bang nucleosynthesis \cite{Lin:2011gj}, while the orange and yellow regions are excluded by constraints on the observed ellipticity of certain Dark Matter halos \cite{Lin:2011gj}}%
\label{fig:ratio}%
\end{figure}

Fig.~\ref{fig:ratio} explores the values of the mediator mass and fermionic coupling (to quarks, $g_q$) relevant for the ratios $g_n/m_\phi^2$ that will be tested by NEWS, for $m_X=1$ GeV. The region shaded in grey is ruled out by the requirement of the lifetime of the mediator not exceeding the epoch of Big Bang nucleosynthesis \cite{Lin:2011gj}; the regions shaded in yellow and orange, instead, correspond to mediator masses excluded by constraints on the observed ellipticity of Dark Matter halos, for scalar and vector mediators, as indicated  \cite{Lin:2011gj}. We also indicate, with the same line conventions, the projected sensitivities of NEWS-Ne and of SuperCDMS Si HV for the particular value of the particle dark matter mass chosen for the plot.

The figure illustrates that there exists a large and natural theory parameter space compatible with direct detection cross sections that will be probed by the NEWS apparatus. Similar conclusions apply for different values of the DM particle mass $m_X$.

\section{Discussion and Conclusions}
\label{sec:conc}
We have evaluated some aspects of the science impact of the NEWS apparatus \cite{Gerbier:2014jwa} in probing certain specific low-mass, GeV-scale Dark Matter candidates. We focused on two classes of theories: millicharged Dark Matter, and light Dark Matter coupled to the Standard Model through a light scalar or vector mediator (of course there exists a broad array of alternative possibilities to the two examples we decide to study here, see e.g. \cite{Izaguirre:2015yja} and references therein). 

For both classes of theories, we showed that NEWS has unique potential in testing viable and otherwise unexplored parameter space regions (especially at very low masses, outperforming other experimental setups such as SuperCDMS Si HV for masses below around 0.3 GeV), and we have detailed the relevant physical properties of such regions: for millicharged Dark Matter, the effective charge is between a billionth and a millionth the unit electric charge $e$, while for the light mediator scenario the ratio of the coupling of the mediator to the nucleons over the mediator mass squared in GeV units ranges between $10^{-6}$ and $10^{-2}$. NEWS will be unique in probing the parameter space regions of very low mass (lighter than around 0.3 GeV) which are beyond the sensitivity of planned solid-state detectors.

\section*{Acknowledgments}
\noindent  I am grateful to Gilles Gerbier for introducing me to the NEWS detector concept, for related illuminating discussions, and for providing the updated sensitivity reach shown in the figures; I am also grateful to Ianis Giomataris for feedback on the manuscript. SP is partly supported by the US Department of Energy under contract DE-FG02-04ER41268. 

%\clearpage


\begin{thebibliography}{300}

\bibitem{Giomataris:2008ap} 
  I.~Giomataris, I.~Irastorza, I.~Savvidis, S.~Andriamonje, S.~Aune, M.~Chapelier, P.~Charvin and P.~Colas {\it et al.},
  %``A Novel large-volume Spherical Detector with Proportional Amplification read-out,''
  JINST {\bf 3}, P09007 (2008)
  [arXiv:0807.2802 [physics.ins-det]]; Website: {\tt website.address.here}
  
\bibitem{Dastgheibi-Fard:2014maa} 
  A.~Dastgheibi-Fard, I.~Giomataris, G.~Gerbierb, J.~Derree, M.~Gros, P.~Magnier, D.~Jourde and E.~.Bougamont {\it et al.},
  %``Background optimization for a new spherical gas detector for very light WIMP detection,''
  PoS TIPP {\bf 2014}, 375 (2014)
  [arXiv:1412.0161 [astro-ph.IM]].

\bibitem{Gerbier:2014jwa} 
  G.~Gerbier, I.~Giomataris, P.~Magnier, A.~Dastgheibi, M.~Gros, D.~Jourde, E.~Bougamont and X.~F.~Navick {\it et al.},
  %``NEWS : a new spherical gas detector for very low mass WIMP detection,''
  arXiv:1401.7902 [astro-ph.IM].
  %%CITATION = ARXIV:1401.7902;%%
 
\bibitem{Cushman:2013zza} 
  P.~Cushman, C.~Galbiati, D.~N.~McKinsey, H.~Robertson, T.~M.~P.~Tait, D.~Bauer, A.~Borgland and B.~Cabrera {\it et al.},
  %``Working Group Report: WIMP Dark Matter Direct Detection,''
  arXiv:1310.8327 [hep-ex].
  %%CITATION = ARXIV:1310.8327;%%
  
  \bibitem{supercdms}
 http://indico.cern.ch/event/276476/session/8/contribution/20/attachments/501916/693116/SuperCDMS\_Aspen2014.pdf
 
 \bibitem{Aguilar-Arevalo:2013uua} 
  A.~A.~Aguilar-Arevalo {\it et al.} [DAMIC Collaboration],
  %``DAMIC: a novel dark matter experiment,''
  arXiv:1310.6688 [astro-ph.IM].
  
  \bibitem{GerbierPC}
  G.~Gerbier, private communication; see also {\tt https://www.snolab.ca/news-projects/}
  
  
  \bibitem{Goldberg:1986nk} 
  H.~Goldberg and L.~J.~Hall,
  %``A New Candidate for Dark Matter,''
  Phys.\ Lett.\ B {\bf 174}, 151 (1986).
  
  \bibitem{DeRujula:1989fe} 
  A.~De Rujula, S.~L.~Glashow and U.~Sarid,
  %``Charged Dark Matter,''
  Nucl.\ Phys.\ B {\bf 333}, 173 (1990).

\bibitem{Dimopoulos:1989hk} 
  S.~Dimopoulos, D.~Eichler, R.~Esmailzadeh and G.~D.~Starkman,
  %``Getting a Charge Out of Dark Matter,''
  Phys.\ Rev.\ D {\bf 41}, 2388 (1990).


\bibitem{Davidson:1993sj} 
  S.~Davidson and M.~E.~Peskin,
  %``Astrophysical bounds on millicharged particles in models with a paraphoton,''
  Phys.\ Rev.\ D {\bf 49}, 2114 (1994)
  [hep-ph/9310288].

\bibitem{Davidson:2000hf} 
  S.~Davidson, S.~Hannestad and G.~Raffelt,
  %``Updated bounds on millicharged particles,''
  JHEP {\bf 0005}, 003 (2000)
  [hep-ph/0001179].


\bibitem{Feldman:2007wj} 
  D.~Feldman, Z.~Liu and P.~Nath,
  %``The Stueckelberg Z-prime Extension with Kinetic Mixing and Milli-Charged Dark Matter From the Hidden Sector,''
  Phys.\ Rev.\ D {\bf 75}, 115001 (2007)
  [hep-ph/0702123 [HEP-PH]].

\bibitem{Holdom:1985ag} 
  B.~Holdom,
  %``Two U(1)'s and Epsilon Charge Shifts,''
  Phys.\ Lett.\ B {\bf 166}, 196 (1986).

\bibitem{Billard:2013qya} 
  J.~Billard, L.~Strigari and E.~Figueroa-Feliciano,
  %``Implication of neutrino backgrounds on the reach of next generation dark matter direct detection experiments,''
  Phys.\ Rev.\ D {\bf 89}, no. 2, 023524 (2014)
  [arXiv:1307.5458 [hep-ph]].
 
\bibitem{McDermott:2010pa} 
  S.~D.~McDermott, H.~B.~Yu and K.~M.~Zurek,
  %``Turning off the Lights: How Dark is Dark Matter?,''
  Phys.\ Rev.\ D {\bf 83}, 063509 (2011)
  [arXiv:1011.2907 [hep-ph]].

\bibitem{Chuzhoy:2008zy} 
  L.~Chuzhoy and E.~W.~Kolb,
  %``Reopening the window on charged dark matter,''
  JCAP {\bf 0907}, 014 (2009)
  [arXiv:0809.0436 [astro-ph]].

\bibitem{Foot:2010yz} 
  R.~Foot,
  %``Do magnetic fields prevent mirror particles from entering the galactic disk?,''
  Phys.\ Lett.\ B {\bf 699}, 230 (2011)
  [arXiv:1011.5078 [hep-ph]].

 \bibitem{Sander:2012nia} 
  J.~Sander, Z.~Ahmed, A.~J.~Anderson, S.~Arrenberg, D.~Balakishiyeva, R.~B.~Thakur, D.~A.~Bauer and D.~Brandt {\it et al.},
  %``SuperCDMS status from Soudan and plans for SNOLab,''
  AIP Conf.\ Proc.\  {\bf 1534}, 129 (2012).

\bibitem{Lin:2011gj} 
  T.~Lin, H.~B.~Yu and K.~M.~Zurek,
  %``On Symmetric and Asymmetric Light Dark Matter,''
  Phys.\ Rev.\ D {\bf 85}, 063503 (2012)
  [arXiv:1111.0293 [hep-ph]].

\bibitem{Izaguirre:2015yja} 
  E.~Izaguirre, G.~Krnjaic, P.~Schuster and N.~Toro,
  %``Accelerating the Discovery of Light Dark Matter,''
  arXiv:1505.00011 [hep-ph].


\end{thebibliography}
\end{document}